# Geometry Limitations in Indirect Selective Laser Sintering of Alumina

D. Sassaman[1]*, M. Ide†, J. Beaman*, C. Seepersad*, and D. Kovar*

*Center for Additive Manufacturing and Design Innovation and
Walker Department of Mechanical Engineering
The University of Texas at Austin, Austin, TX, 78712

†ExxonMobil Research and Development Company
Annandale, NJ 08801

## Abstract

Ceramics containing open channels with complex geometries can be manufactured by additive manufacturing (AM) and are of great interest in clean energy technologies. However, design limitations and guidelines for manufacturing these architectures with AM have not yet been established. In this work, we compare previously proposed geometry limitations for polymer selective laser sintering (SLS) to the geometries produced using indirect SLS in alumina. We focus on a subset of model shapes that are simple to produce and measure. We show that these rules provide a starting point for the design and manufacture of ceramic geometries using indirect SLS. However, there are additional considerations for AM of ceramics by indirect SLS that further limit the geometries that can be produced.

## 1. Introduction

Indirect selective laser sintering (SLS) of ceramics uses a sacrificial polymer binder that is mixed with a ceramic powder before processing. Indirect SLS simply replaces the consolidation/forming step in traditional ceramic processing; the same pre- and post-processing steps are required (Figure 1). In the case of indirect SLS, this consolidation step is comprised of two sub-processes; powder layer deposition and laser sintering. The spreading mechanism deposits a thin layer of a ceramic/binder mixture. During laser irradiation, only the binder melts, forming bridges between the ceramic particles. In this paper, to avoid confusion, the term "sintering" is defined as the act of joining particles together during the SLS process, and "densification" is defined as joining the ceramic particles during post-processing.

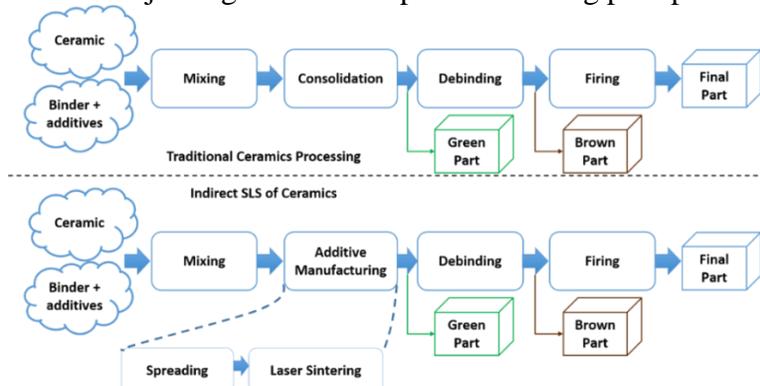

*Figure 1: Top) Traditional ceramic processing, Bottom) Indirect SLS ceramic processing*

---

[1] Corresponding author contact information: doug.sassaman@utexas.edu

Complex geometries produced by *indirect* SLS have been studied in the context of cellular ceramics, metamaterials, flow reactors, and tissue scaffold engineering [1], but rules and limitations for designing them have not been developed. Existing literature primarily focuses on geometrical accuracy of relatively simple shapes [2]–[6]. The most recent and comprehensive work on this topic comes from researchers at KU Leuven [3] and The University of Missouri Rolla [7]. For example, Nolte *et al.* were able to produce holes with relatively complex paths and straight holes with diameters of 1 mm ± 0.12 mm [7]. Nissen *et al.* proposed a methodology for determining geometry limitations in helical glass channels produced by indirect SLS [8]. Allison *et al.* developed geometry limitations, and a method for determining them, for polymer SLS [9]–[11]. Additionally, Allison *et al*. developed a metrology part, from which the geometries in this work have been adapted.

The purpose of this paper is twofold; 1) test geometry limitations on indirect SLS which were originally developed for polymer SLS, and 2) catalog phenomena and limitations associated with indirect SLS (some of which may not be present in polymer SLS).

## 2. Materials and Methods

### *Materials*

This work uses a blended alumina/nylon powder system as a testbed, which was adapted from Deckers *et al.* [12]. A mixture of powders consisting of 78 wt.% alumina (Almatis A16 SG, $d_{50}$=0.3µm) with 22 wt.% PA12 (ALM PA650 $d_{50}$=58µm) was dry-mixed in a high-shear blender (Chulux QF-TB159008) for 10 minutes. The blended mixture was then sieved through a 250 µm mesh to produce the powders shown in Figure 2 and Figure 3.

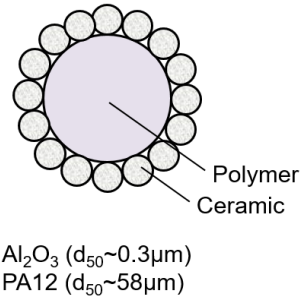

$Al_2O_3$ ($d_{50}$~0.3µm)
PA12 ($d_{50}$~58µm)

*Figure 2: Schematic representation of morphology of the powder used in this work consisting of alumina and nylon (PA12)*

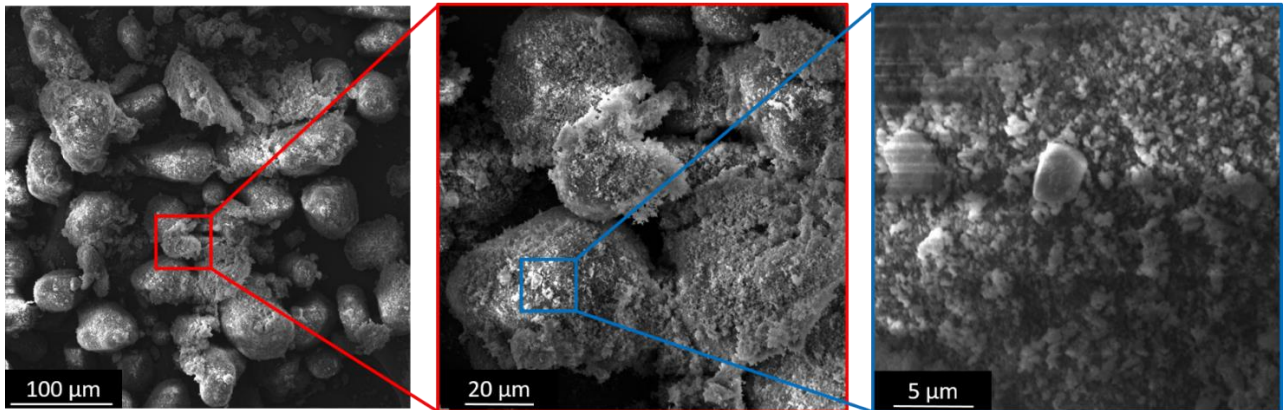

*Figure 3: Scanning electron micrographs of blended alumina and nylon powders used in this work at a range of magnifications.*

## Methods: SLS Machine

A custom SLS machine at The University of Texas at Austin, known as the Laser Additive Manufacturing Pilot System (LAMPS) was used for this study. LAMPS has extensive sensing and control capabilities built into open-hardware and open-software architectures, which make it ideal for developing suitable process parameters for experimental materials [13].

The parameters used for SLS were selected empirically with the help of visual and thermal imaging to ensure that degradation of the binder did not occur and that part curl was minimized [14]. The final parameters that were used are as follows:

- Laser power: 4 - 10 W
- Laser scan speed: 200 - 1000 mm/s
- Layer thickness: 100 µm
- Beam (hatch) spacing: 275 µm
- Spot size[2]: $1/e^2 = 730$ µm, full-width half-max (FWHM) = 580 µm
- Powder bed temperature[3]: 179 - 191°C

## Methods: Debinding and Densification

After the green parts were removed from the SLS machine, loose powder was removed with a focused stream of dry air at a pressure of 680 kPa. Thermal debinding was performed in a tube furnace, with flowing dry air (~0.1 standard litre per minute) used to carry away the decomposition products. Each sample was placed in an alumina crucible on top of a thin layer of coarse-grained alumina (*Goodfellow GF18024511, $d_{50}=45$ µm*) to avoid sticking of the part to the crucible. The crucibles were loosely covered to reduce direct contact with the flowing air. The parts were heated at 9°C/hour to 600°C, and then cooled to room-temperature naturally. After debinding, parts were transferred to a box furnace for final densification. Temperature was increased at a rate of 10°C/minute, and then held at 1600°C for 1 hour, before returning to room temperature at a rate of 10°C/minute. These heating and cooling rates were selected to minimize distortion of the part but were not optimized for overall processing time.

## Methods: Determining Geometry Limitations

After selecting a material system and processing route, a factorial-style approach that was originally developed by Allison *et al.* was used to determine geometry limitations for samples containing through-thickness channels with two simple cross-sectional shapes: holes and slots (Figure 4 and Table 1). Laser power, and feature orientation were varied for each of these shapes. There were two sub-methods for this procedure; 1) <u>Qualitative</u>: Visual inspection was used to determine if a feature has resolved, and 2) <u>Quantitative:</u> Pin gages were to measure dimensions of the holes and slots.

The qualitative approach was used to assess if each feature (*e.g.*, hole) resolved properly. In this work, a channel resolved if, by visual inspection, light passed through the channel. The quantitative approach used analysis of variance (ANOVA), with a confidence interval of 95%, to determine *which* factors significantly affected the outcome. The factors studied here were laser power and feature orientation. After performing the ANOVA analysis, *post-hoc* testing was used

---

[2] Measured via Ophir NanoScan
[3] Monitored with a FLIR A6701 mid-wave infrared camera, which was used to control quartz lamps that heated the powder bed via PID control

to quantify the degree to which the significant input factors affected the output. For example, *if* laser power significantly affected the measured hole diameter, *by how much*? This procedure is known as a multiple comparisons test or a Tukey-Kramer test [15]. An additional metric, the *average mean difference,* is included for clarity for some tests. The average mean difference determines the average amount that the feature of interest (*e.g.,* hole diameter) is changed for a given type of input factor change (*e.g.*, changing feature orientation from vertical to horizontal). The geometries shown in Figure 4 were produced in duplicate for each setting. So, for example, the part labeled "Large Holes" was produced twice vertically and twice horizontally for each laser power setting.

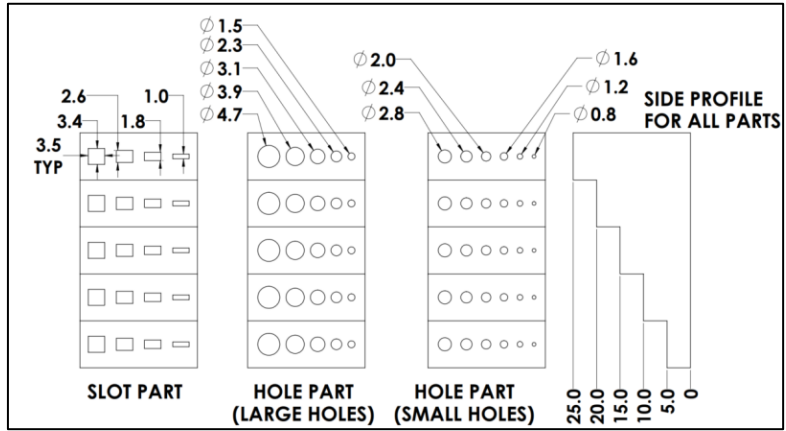

*Figure 4: Hole and slot channels. Hole diameters ranged from 0.8 mm - 4.7 mm with depths of 5 mm - 25 mm Slot widths ranged from 1 mm - 3.4 mm with depths of 5 mm- 25 mm. Slot height was held constant at 3.5 mm.*

*Table 1: Features and input factors for holes and slots*

| Feature | Feature Range(s) & Part Specs. | No. of Replicates (per feature per setting) | Factor(s) & Setting(s) | | | |
|---|---|---|---|---|---|---|
| | | | Laser Power [W] | Scan Speed [mm/s] | Orientation | Raster |
| **Hole** | 0.8 - 4.7 mm hole diameter | 2 | 4.7, 6.5, 8.5, 9.1, 11.7, 14 | 1000 | Horizontal, Vertical | 275 μm hatch spacing |
| **Slot** | 1.56 - 3.45 mm hydraulic diameter | 2 | 4.7, 6.5, 8.5 | 1000 | Horizontal, Vertical | 275 μm hatch spacing |

The quantitative comparisons presented for this study were measured using pin gages inserted into the holes or slots. Holes and slots measured using this method are characterized by their hydraulic diameters, $D_{hyd,hole}$ or $D_{hyd,slot}$, so that comparisons can be made between them:

$$D_{hyd,hole} = D_{hole} \quad (1)$$

$$D_{hyd,slot} = \frac{4WH}{2W + 2H} \quad (2)$$

$W = slot\ width, H = slot\ height, D_{hole} = hole\ diameter$

The ratio of the nominal hydraulic diameter to the depth is defined from

$$Nominal \frac{Hydraulic\ Diameter\ [mm]}{depth\ [mm]} \equiv \frac{D}{d} \qquad (3)$$

where nominal refers to the specified build dimension rather than the actual measured dimension.

The first attempt to quantitatively measure the hole and slot hydraulic diameters was adopted from Allison *et al*. Unfortunately, this approach, which used an optical flatbed scanner to measure holes sizes, was determined to be inadequate for the present work *after* some parts had already been densified. Therefore, a few comparisons with green parts are not possible because the final methodology for quantifying geometrical features had not been developed when these parts were densified.

## 3. Results

### A: Test of Geometry Limitations on Indirect SLS which were Developed for Polymer SLS

#### Qualitative Results

There are two questions to be addressed in this section; I) are geometrical features in parts produced by indirect SLS less likely to resolve than in polymer parts, and II) Are the effects of processing parameters on geometrical limitations for SLS of polymer parts the same as for ceramic parts produced by indirect SLS.

To address question (I), the "Large Holes" part from Figure 4 is used. Figure 5 shows an example comparison between a nylon part produced by SLS by Alison *et al.* and a densified alumina part produced for this work. Figure 6 (A – B) addresses the differences in likelihood of resolution of holes in nylon parts and in alumina parts. Each cell contains a number that is the percentage of channels with that diameter and depth that resolved correctly. The boxes are shaded from dark-grey, which indicates that all of the parts resolved correctly, to white, which indicates that none of the parts resolved correctly.

Figure 6 shows that for both materials, channels with large hole diameters and shallow depths are more likely to resolve properly than channels with small diameters and large depths. None of the channels in the ceramic part collapsed during debinding and densification, so the resolution tables are identical between the green and final parts. Although the range of features studied is different between the material systems, within the overlapping region (*e.g.*, hole depth 5-10 mm, hole diameter 0.8-2 mm), the nylon parts show a higher likelihood of resolving a feature in every instance.

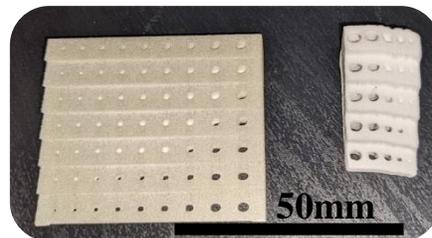

*Figure 5: Comparison of parts containing holes part in a nylon part from Allison et al. (Left) to holes in a **densified** alumina part from this work (Right)*

A) **Holes in Nylon Parts**

| Hole Diameter [mm] \ Hole Depth [mm] | 1 | 2.5 | 4 | 5.5 | 7 | 8.5 | 10 |
|---|---|---|---|---|---|---|---|
| 0.8 | 0.69 | 0.41 | 0.23 | 0.05 | 0.04 | 0.04 | 0.04 |
| 1 | 0.76 | 0.57 | 0.35 | 0.24 | 0.13 | 0.07 | 0.06 |
| 1.2 | 0.85 | 0.70 | 0.49 | 0.35 | 0.22 | 0.16 | 0.08 |
| 1.4 | 0.92 | 0.78 | 0.60 | 0.43 | 0.36 | 0.26 | 0.17 |
| 1.6 | 0.94 | 0.86 | 0.69 | 0.55 | 0.43 | 0.37 | 0.32 |
| 1.8 | 0.95 | 0.88 | 0.80 | 0.60 | 0.51 | 0.43 | 0.4 |
| 2 | 0.96 | 0.91 | 0.83 | 0.68 | 0.58 | 0.5 | 0.5 |

B) **Holes in Alumina Parts**

| Hole Diameter [mm] \ Hole Depth [mm] | 5 | 10 | 15 | 20 | 25 |
|---|---|---|---|---|---|
| 0.8 | 0.00 | 0.00 | 0.00 | 0.00 | 0.00 |
| 1.2 | 0.17 | 0.00 | 0.00 | 0.00 | 0.00 |
| 1.5 | 0.08 | 0.00 | 0.00 | 0.00 | 0.00 |
| 1.6 | 0.17 | 0.00 | 0.00 | 0.00 | 0.00 |
| 2 | 0.58 | 0.25 | 0.00 | 0.00 | 0.08 |
| 2.3 | 0.58 | 0.33 | 0.17 | 0.17 | 0.08 |
| 2.4 | 0.75 | 0.83 | 0.50 | 0.25 | 0.25 |
| 2.8 | 0.75 | 0.92 | 0.67 | 0.42 | 0.17 |
| 3.1 | 0.83 | 0.58 | 0.50 | 0.42 | 0.33 |
| 3.9 | 1.00 | 0.83 | 0.67 | 0.50 | 0.50 |
| 4.7 | 1.00 | 0.92 | 0.83 | 0.83 | 0.75 |

C) **Slots in Alumina Parts**

| Slot Hydraulic Diameter [mm] \ Slot Depth [mm] | 5 | 10 | 15 | 20 | 25 |
|---|---|---|---|---|---|
| 1.56 | 0.50 | 0.33 | 0.17 | 0.00 | 0.00 |
| 2.38 | 1.00 | 0.92 | 0.83 | 0.58 | 0.50 |
| 2.98 | 1.00 | 1.00 | 1.00 | 0.92 | 1.00 |
| 3.45 | 1.00 | 1.00 | 1.00 | 1.00 | 0.92 |

D)

| | Proportion Resolved | | | |
|---|---|---|---|---|
| Diameter [mm] → | 0.8 | 2 (2.38 for slots) | 0.8 | 2 (2.38 for slots) |
| Depth [mm] → | 5 (5.5 for Nylon) | 5 (5.5 for Nylon) | 10 | 10 |
| Nylon Holes | 0.05 | 0.68 | 0.04 | 0.50 |
| Alumina Holes | 0.00 | 0.58 | 0.00 | 0.25 |
| Alumina Slots | N/A | 0.92 | N/A | 1.00 |

*Figure 6: Percentage of parts (indicated by number in each cell of the table) where features resolved for holes in nylon parts (A), reproduced from Allison et al. [9], holes in alumina (B), and slots in alumina (C). Comparison of the results for all of the material systems (D).*

Question (II) is addressed by comparing the results for holes and slots in alumina parts, which is shown in Figure 6 B and C, respectively. It is apparent that slots resolve to smaller diameters and larger depths than holes. For example, 50% of holes resolved for a 3 mm hydraulic diameter and 15 mm depth, but 100% of the slots resolved with the same hydraulic diameter and depth. This trend, that slots are more likely to resolve than holes, was also observed by Allison *et al.* for nylon parts, which indicates that the qualitative trends in feature resolution using indirect SLS of ceramics are similar to polymer SLS.

In summary, it is likely that the qualitative method developed for polymer SLS can also be applied to indirect SLS of alumina. For most features, the likelihood of resolution is the same order of magnitude for both systems, and the qualitative trends are similar.

### *Quantitative Results*

This section will address similar questions to the qualitative section above, but now from a quantitative perspective. To address question (I), the accuracy relative to the specified dimension of holes and slots in alumina is compared to holes in nylon of similar specified dimension. As will be shown later, in some cases, the error in hydraulic diameters of the ceramic parts in their green state rivaled that of the nylon parts. However, shrinkage during debinding and densification increased this error significantly. The results from the densified alumina parts are presented in this section.

Figure 7 shows that geometries produced by SLS in nylon are ~5 – 10× more accurate than when produced by indirect SLS in alumina. In both material systems, slots are more accurate than holes on average. However, the deviation from specified dimensions is much larger in the alumina system (~20 - 80%) than in the nylon system (2 - 22%) for both holes and slots. It is likely that the larger error of the alumina system can be attributed to the heterogeneity of both agglomerate size and the distribution of materials in the powders used for indirect SLS.

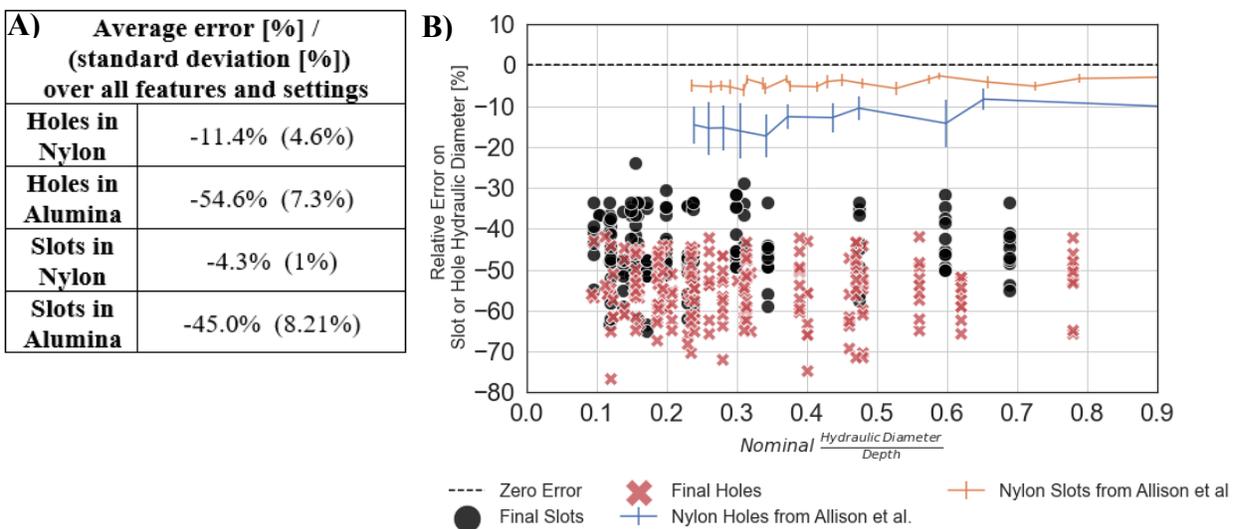

Figure 7: Average error and standard deviation of hydraulic diameter for slots and holes for densified alumina and nylon [Taken from 9] over all features and settings (A). Comparison between holes and slots for densified alumina parts and nylon parts (B).

Question (II) is addressed by comparing a subset of similarly sized nylon and alumina channels (Figure 8). The accuracy of the hole or slot is assessed by comparing the specified diameter to the diameter of that channel measured using a pin gauge. For holes in nylon parts, Allison *et al.* observed a decrease in the relative error in the diameter with decreasing hole depth. For slots in nylon parts, they found the accuracy to be relatively insensitive to the depth of the slot. In contrast, the relative errors for holes and slots in alumina parts produced for this study exhibit such large variances that they seem to be insensitive to the depth for the range tested. The data confirm that there are fundamental differences in the factors that limit the geometry for indirect SLS compared to conventional SLS and that the quantitative trends observed in nylon parts are not replicated in indirect SLS.

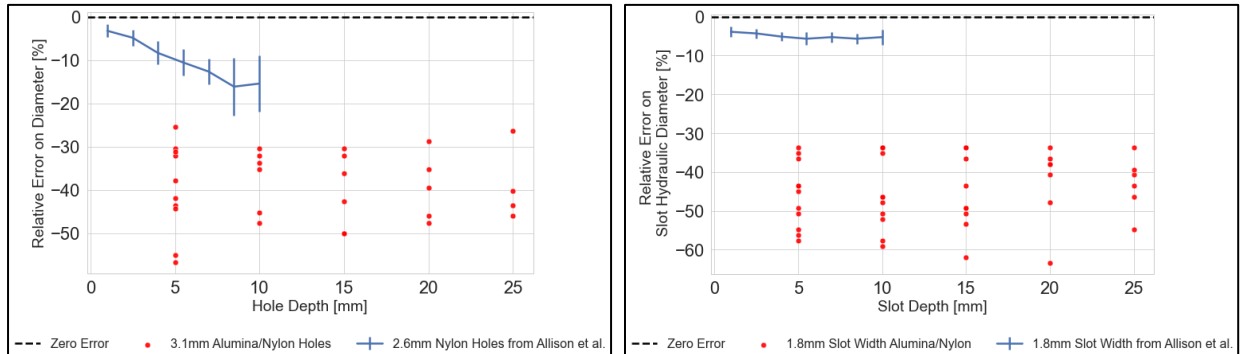

*Figure 8: Left) Hole accuracy as a function of hole depth for alumina/nylon in green state and nylon-only holes part from Allison et al.[10]. Right) Slot accuracy as a function of slot depth.*

In summary, the data indicate that, the ability to produce a hole or slot that penetrates through the thickness of the part are similar for the two material systems. However, where the material systems differ significantly is in the measured *accuracy* of the holes and slots. The measured accuracy of these features is significantly poorer in alumina parts produced by indirect SLS compared to SLS of polymers.

### B: Catalog of Phenomena and Limitations Associated with Indirect SLS

The goal of this section is document phenomena that influence feature accuracy and that are specific to the process of producing alumina channels using indirect SLS. These phenomena were investigated *via* the quantitative procedure described earlier; ANOVA and *post-hoc* testing for each hole and slot at each depth was performed. Full results are presented in the appendix, but the pertinent information is included here. This section is separated into two parts; A) comparison of holes *vs.* slots, and B) phenomena associated with debinding and densification.

#### Holes vs Slots

To compare the phenomena associated with holes and slots, the relative error in hydraulic diameter versus the *D/d* ratio (Figure 9 A-B) is used in conjunction with the average mean differences (Figure 9 C). Mean differences are the relative changes in output, *e.g.*, hole diameter, between *each* input settings whereas, average mean differences are the average of all mean differences for a given type of feature. Mean difference therefore can be used to explain how a particular feature (*e.g.*, hole with depth of 10 mm and diameter of 2 mm) would respond to a given change in laser power. On the other hand, the average mean difference can be used to explain how a given change in laser power affects holes of *all* sizes. The only input factor which can be directly compared is feature orientation since this parameter was systematically varied in both this study and the study by Allison *et al*. All of the input factors for both the Allison *et al.* work and this work are provided in Figure 9 C because their relative magnitudes are important. The build parameters that were varied include: 1) the material (for nylon parts only) used for the build was Nylon 12, Nylon 12 reinforced with glass beads, and fire retardant Nylon 11, 2) the feature orientation was either with the long axis of channel parallel or perpendicular to the laser beam 3) the location on the build platform (for the nylon parts only), and 4) the laser power used to produce the part (for the alumina samples only).

Figure 9 A-B show the effects of specific build parameters on the relative error in diameter versus the *D/d* ratio. The color and shape of the data points specify the input factor (*e.g.*, laser power). Figure 9 A does not show a significant influence of feature orientation on the relative error in hole diameter. In contrast, the relative error in hole diameter does appear to increase with laser power. However, a careful analysis using the full ANOVA results presented in the appendix showed that laser power is not a significant factor for all holes and hole depths. In contrast to the errors in hole dimensions, Figure 9 B suggests that the errors in the slot dimensions are somewhat agnostic to input factors, especially orientation; similar results were observed previously in nylon-only parts. [10], [11].

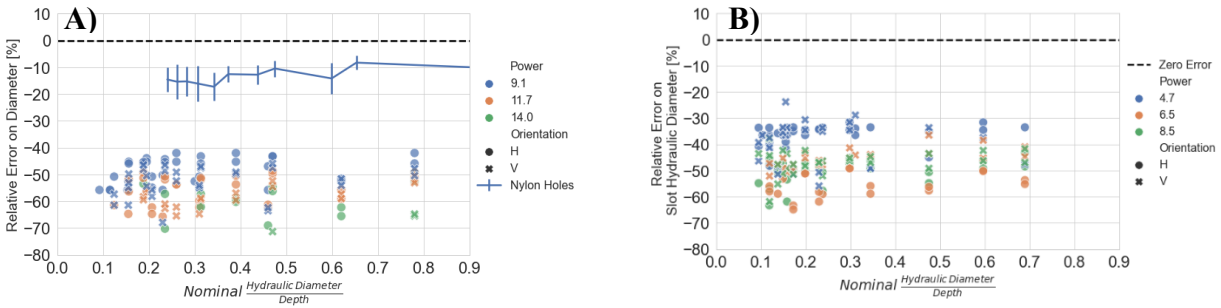

### C) Average mean difference [mm] / Percent Significant

|  | Material | Orientation | Location | Laser Power |
|---|---|---|---|---|
| Nylon Holes | 0.26 / 100% | 0.32 / 100% | 0.13 / 29% | N/A |
| Alumina Holes | N/A | 0.25 / 85% | N/A | 0.31 / |
| Nylon Slots | 0.21 / 100% | 0.09 / 14% | 0.04 / 57% | N/A |
| Alumina Slots | N/A | -0.17 / 33% | N/A | 0.10 / 83% |

*Figure 9: Relative error in diameter versus the D/d ratio for holes (A), and slots (B). In (A), the solid blue line is for nylon holes from Allison et al.[10]. Scatter points are for densified alumina parts produced for this study. Build variables for alumina parts are laser power and feature orientation. (C) Shows a comparison of average mean differences between nylon [11] and densified ceramic geometries. Percent significant indicates how often a given factor was significant. For example, orientation was a significant factor in alumina slots for 4 of the 12 instances studied.*

The average mean differences due to changing input factors on each feature are presented in Figure 9. Interestingly, the average mean differences on hydraulic diameter are of the same order of magnitude between material systems. This means, for example, that the difference between slots built in horizontal and vertical orienations is similar between material systems (Δ=0.17mm for alumina, Δ=0.09mm for nylon). Perhaps the most interesting conclusion from this figure is that holes and slots respond differently to input factors, which is seen in both material systems. This indicates that the differences between holes and slots results from the SLS process, rather than from the material systems themselves. The data for slots in alumina parts suggest only a weak dependence on laser power. The negative value in the orientation column indicates that slots built in the vertical orientation tend to be larger than slots built in the horizontal orientation. Note that this is not the case for slots in nylon parts, or for holes in either material system. This table should be used in conjunction with the full ANOVA results in the appendix, which shows that orientation is only a significant factor for a minority of combinations of slot width and depth (33%).

In summary, this section shows that holes and slots in alumina parts respond differently to input factors, but that this was also the case for holes and slots in nylon parts. Therefore, the phenomena cataloged here are not unique to alumina parts produced with indirect SLS, but are driven by the underlying physics of the SLS process.

*Debinding and Densification*

There is some indication from the results for holes presented in Figure 10 A that tuning the process parameters for the indirect SLS system can produce green parts with similar accuracy to their nylon-only counterparts. However, it is possible that the accuracy can be further degraded during debinding and sintering. Deformation during debinding and densification has been well documented during conventional processing of ceramics (See Rahaman [16]). These phenomena are no different for indirect SLS, aside from the fact that slightly more binder is used in indirect SLS than traditional ceramics processes, such as powder injection molding (PIM). However, Figure 10 E shows that the errors are reduced in parts after densification. The standard deviation of error in hole diameter (over all hole sizes and input factors) for holes in the green state is 11.4%, while it is 7.3% for holes in densified parts. The mechanisms for this reduction in error will be the focus of a later study, but preliminary results are presented here.

Figure 10 (C-D) shows that there is a slight decrease in the relative shrinkage of the hole with increasing hole diameter, but no statistical difference with feature orientation. The plots also show that the relative shrinkage decreases with increasing laser power. Higher laser powers may result in larger melt zones that facilitate particle rearrangement during the sintering process. It is likely that this dependency is affected by the heterogeneous nature of the mixed alumina/nylon system. Further testing needs to be performed to confirm the cause of this phenomena.

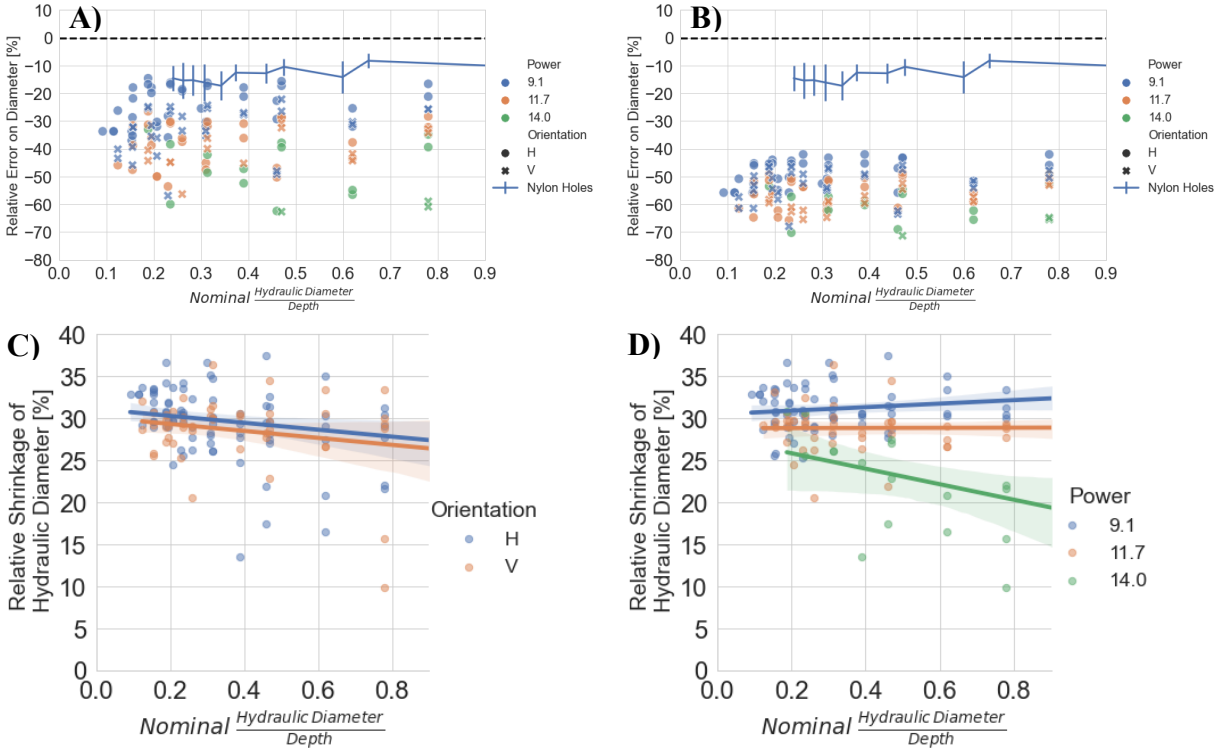

Figure 10: Relative error in diameter versus the D/d ratio for holes in green state (A), and densified state (B). Relative shrinkage in diameter for holes varying orientation (C), and with varying laser power (D). Comparison of the average mean differences between green and final holes (E).

## Conclusions

We have shown that the previously proposed geometry limitations for polymer SLS provide a starting place for the design and manufacture of ceramic geometries using indirect SLS. For example, green parts produced by indirect SLS show some similar dependencies on process input factors. However, the dimensional errors in alumina parts are considerably larger than those measured previously in nylon parts produced by SLS. In some cases, it is possible to improve dimensional accuracy of green parts to a level close to that of polymer parts by specifically tuning build parameters for indirect SLS. Even in these cases, binder removal and densification was shown to result in further degradation in the dimensional accuracy of alumina parts. Some trends in dimensional errors observed in direct SLS of nylon also occur in indirect SLS of alumina; for example, the accuracy of holes is more dependent on feature orientation than is the accuracy of slots. There are also process-specific dependencies which warrant further research such as the dependence of part shrinkage on the laser power used to produce the green part.

# Acknowledgements

This work was supported by ExxonMobil through its membership in The University of Texas at Austin Energy Institute.

# Appendix A: Pictures of Parts

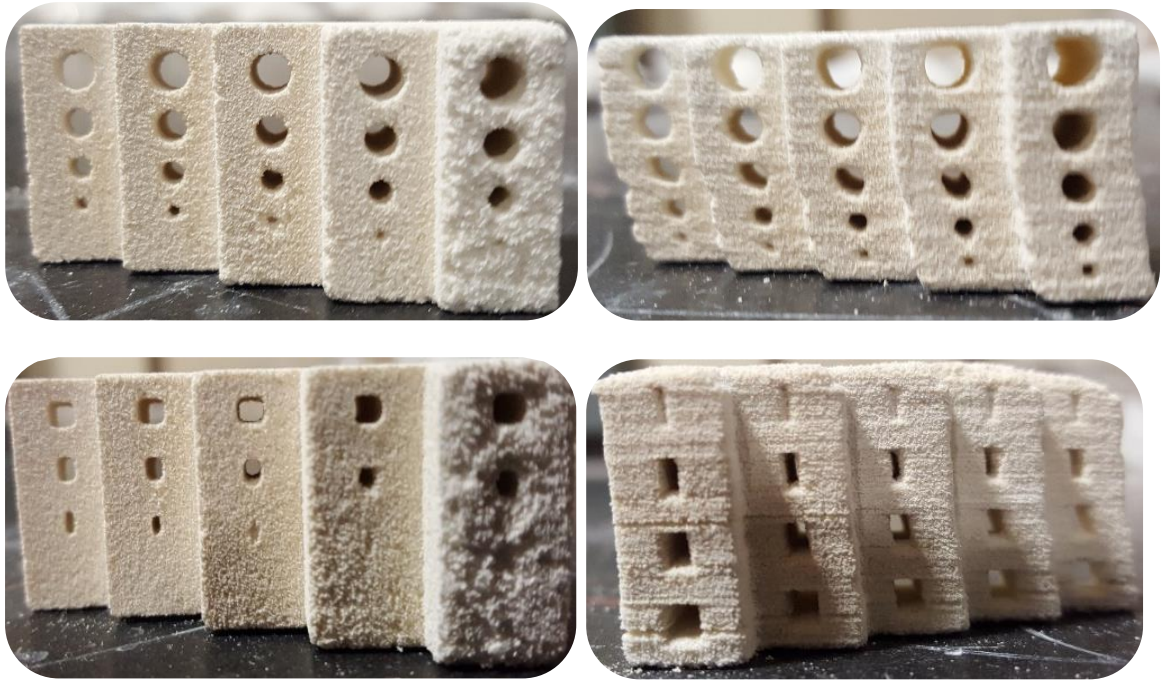

*Figure 11: Holes and slots in alumina/nylon in the green state. From top left, clockwise: A) Holes, laser power = 9.1 W, vertical feature orientation, B) Holes, laser power = 9.1 W, horizontal feature orientation, C) Slots, laser power = 8.5 W, vertical feature orientation, D) Slots, laser power = 8.5 W, horizontal feature orientation*

# Appendix B: Full ANOVA Results

The following tables show significant effects and the mean differences associated with each effect. The numbers inside of the parentheses show, on average, how much that factor affected hole diameter (in mm). For example, for a hole diameter of 3.1 mm and depth of 5 mm, each change in power level decreased hole diameter by ~0.37 mm. Cells listed as N/A are instances which were not studied because their resolution likelihood was so small.

*Table 2: ANOVA and mean differences for **holes** in **alumina-nylon** in the **green state**. Letters indicate which factors were significant; P = Power; O = Orientation, P*O = interaction.*

| | Significant effects, (mean differences [mm]) | | | | |
|---|---|---|---|---|---|
| Hole Depth [mm] → <br> Hole Diameter [mm] ↓ | 5 | 10 | 15 | 20 | 25 |
| 1.5 | N/A | N/A | N/A | N/A | N/A |
| 2.3 | N/A | N/A | N/A | N/A | N/A |
| 3.1 | P(0.37) | N/A | N/A | N/A | N/A |
| 3.9 | P(0.51), O(0.43), P*O | N/A | P(0.84), O(0.62) | N/A | N/A |
| 4.7 | P(0.74), O(0.36), P*O | P(0.71), O(0.55), P*O | P(0.65), O(0.36) | P(0.78), O(0.55), P*O | P(0.72), O(0.54) |

*Table 3: ANOVA and mean differences for **holes in a alumina** after firing.*
*Letters indicate which factors were significant; P = Power; O = Orientation, P\*O = interaction.*

| Hole Depth [mm] → <br> Hole Diameter [mm] ↓ | Significant effects, (mean differences [mm]) | | | | |
|---|---|---|---|---|---|
| | 5 | 10 | 15 | 20 | 25 |
| 0.8 | N/A | N/A | N/A | N/A | N/A |
| 1.2 | N/A | N/A | N/A | N/A | N/A |
| 1.5 | N/A | N/A | N/A | N/A | N/A |
| 1.6 | N/A | N/A | N/A | N/A | N/A |
| 2.0 | N/A | N/A | N/A | N/A | N/A |
| 2.3 | N/A | N/A | N/A | N/A | N/A |
| 2.4 | P(0.23),O(0.22) | P(0.20),O(0.24) | N/A | N/A | N/A |
| 2.8 | P(0.25) | P(0.22),O(0.08) | | | |
| 3.1 | P(0.17) | N/A | N/A | N/A | N/A |
| 3.9 | P(0.22), O(0.29), P*O | P(0.34), O(0.23) | P(0.45), O(0.34) | N/A | N/A |
| 4.7 | P(0.16), O(0.14) | P(0.43), O(0.37), P*O | P(0.44), O(0.24), P*O | P(0.49), O(0.31), P*O | P(0.38), O(0.24), P*O |

*Table 4: ANOVA and mean differences for **slots in alumina** after firing.*
*Letters indicate which factors were significant; P = Power; O = Orientation, P\*O = interaction effects.*

| Slot Depth [mm] → <br> Slot Hydraulic Diameter [mm] ↓ | Significant effects, (mean differences [mm]) | | | | |
|---|---|---|---|---|---|
| | 5 | 10 | 15 | 20 | 25 |
| 1.56 | N/A | N/A | N/A | N/A | N/A |
| 2.38 | O(-0.21) | P(0.16), O(-0.1), P*O | P(0.16) | P(0.09) | N/A |
| 2.98 | P(0.0), P*O | P(0.17) | P(0.17), O(-0.13) | P(0.17) | NONE |
| 3.45 | O(-0.22) | P(-0.33), P*O | P(0.06), P*O | P(0.12), P*O | N/A |